# Resistive state of quasi-one-dimensional superconductors: fluctuations vs. sample inhomogeneity

M. Zgirski and K. Yu. Arutyunov

*University of Jyväskylä, Department of Physics, PB 35,*

*40014 Jyväskylä, Finland*

**Abstract**

The shape of experimentally observed $R(T)$ transition of thin superconducting wires is analyzed. Broadening of the transition in quasi-1-dimensional superconducting channels is typically associated with phase slip mechanism provided by thermal or quantum fluctuations. It is shown that consideration of inevitable geometrical inhomogeneity and finite dimensions of real samples studied in experiments is of primary importance for interpretation of results. The analysis is based on experimental fact that for many superconducting materials the critical temperature is a function of characteristic dimension of a low-dimensional system: film thickness or wire cross section.

## 1. Introduction

In experiment measured superconducting resistive phase transition $R(T)$ always has a finite width $\Delta T_c$. One might naively expect that for sufficiently pure (chemically homogeneous) sample the transition should be infinitely narrow. It is not the case. The shape of $R(T)$ dependence is governed by fluctuations, being determined by the system dimensionality set by superconducting coherence length $\xi$. In particular limit of quasi-1-dimensional (1D) channels theoretical description is based on concept of fluctuation-governed phase slips activation. At temperature $T$ and sufficiently small measuring current $I$ the finite voltage $V$ of a wire of length $L$ and cross section $\sigma$ ($\sigma^{1/2} < \xi << L$) is:

$$V(T,I) = \Omega(T,I,L)\exp\left\{-\frac{\Delta F}{\varepsilon}\right\}, \qquad (1)$$

where $\Delta F(\sigma)$ is the energy barrier between two potential minima of the system's free energy in phase space. If thermal fluctuations solely determine the behavior of the system [1,2], then $\varepsilon = k_B T$. If quantum fluctuations are important, then the characteristic energy $\varepsilon$ is a more complicated function of sample parameters and temperature [3]. Contribution of the pre-factor $\Omega(T,I,L)$ is negligible compared to strong exponential dependence.

Naturally, theoretical models consider a case of ideal 1D channel where the critical temperature (energy gap) is constant all over the sample. However, even excluding a trivial case of chemically or structurally inhomogeneous systems, one cannot eliminate the contribution of the finite size dependence of the critical temperature. It is a well-known experimental fact that in superconducting films and narrow wires [4] critical temperature might differ from its bulk value. There exist models predicting both reduction [5] and enhancement [6] of $T_c$ in low-dimensional superconductors. In different materials indeed the effect has different sign. The origin of the phenomenon still does not have a commonly accepted explanation. Hereafter we take the size dependence of $T_c$ as granted. We will show that one should be very cautious in fitting experiments on real finite size nanowires with models [1-3] derived for ideal 1D objects. Inevitable size-dependent variations of the local $T_c$ might appear to be the dominating contribution defining the shape of experimentally observed $R(T)$ dependencies.

## 2. Experiment

Classical experiments confirming thermal activation model [1,2] were made in early 70s on perfect tin single crystals (whiskers) with characteristic diameter $\sigma^{1/2} \sim 0.5$ μm [7,8]. Since that time, due to progress in nanofabrication, samples of much smaller dimensions became available. There appeared experiments stating observation of qualitatively new phenomena (quantum fluctuations) in ultra-thin wires with diameter ~10 nm resulting in dramatically broad $R(T)$ dependencies [9,10,4,11]. An example of such transition is presented in Fig. 1 obtained on a single aluminium nanowire by progressive reduction of its diameter [12]. One may ask a reasonable question: how well the existing models derived for ideal 1D superconducting channels can describe experimental reality obtained on far from being perfect real nanowires?

## 3. The model

Let us consider a typical nanowire (thermal or e-beam evaporated lift-off lithography) as relatively clean in a sense that it contains negligible amount of impurities capable of changing critical parameters. Such structures are obviously polycrystalline. If the metal film is at least few tens of nm thick, the level of disorder is not very high: sheet resistance of a typical sample in normal state is of the order of $R_\square \sim 1\Omega$. Only contribution of inevitable geometrical imperfections (cross-section variation along the wire) and finite length will be evaluated.

One can quantify geometrical inhomogeneity of a wire by distribution of its cross sections σ along the length. The distribution can be experimentally obtained from direct measurements using scanning probe microscope (SPM) (Fig.2). Keeping in mind the abovementioned dependence of $T_c$ on wire diameter (Fig. 3, insert), for a quasi-1D structure ($\sigma^{1/2} < \xi$) much longer than the superconducting coherence length $L >> \xi$ it is possible to simulate the $R(T)$ transition as sequence of switchings at different local $T_c$'s of resistors of length $\xi$ connected in series. The effective wire resistance in the region of superconducting transition is given by simple equation:

$$R_{eff}(T) = \int \frac{p(\sigma)\rho(\sigma)L}{\sigma} d\sigma, \qquad (2)$$

where integration should be performed over all cross sections σ with local critical temperature $T_c(\sigma) < T$. Function $p(\sigma)$ characterizes distribution of cross sections (Fig. 2, insert) for particular sample. In the limit of very narrow wires with diameter compared to the electron mean free path $\ell$ normal state resistivity can depend on cross section: $\rho(\sigma)$. If functions $p(\sigma)$ and $\rho(\sigma)$ are known, then the above expression is suitable for numerical calculations. Example of such simulation for typical aluminium nanowire from Fig. 2 with $\sigma^{1/2} \sim 75$ nm is presented in Fig. 3. The electron mean free path $\ell$ for co-deposited film is about 30 nm. For this particular case of 'not-too-narrow' nanowire ($\ell < \sigma^{1/2}$) the size dependence of resistivity can be neglected: $\rho(\sigma) = \rho_{bulk} = const$. If contribution of wider node regions is taken into consideration, then the width of superconducting transition is even larger and being comparable with experimentally obtained $R(T)$ data (Fig. 3). To support our model we have selected the 'worst case scenario' of a short wire with relatively rough surface (Fig. 2 and 3). Experimental $R(T)$ dependencies for smoother and longer samples the are not so that 'bad' as presented in Fig. 3. However, in all cases simulations using thermal phase slip activation model [1,2] for an idealized wire of the same constant cross section result in much narrower $R(T)$ transition.

## 4. Discussion

We have shown that simple consideration of realistic nanowire geometrical inhomogeneity together with size-dependent critical temperature $T_c(\sigma)$ can produce a well defined low-temperature 'foot' on the $R(T)$ dependence. How one can state that broad superconducting transition similar to Fig. 1 indicates presence of a new physical mechanism [4,11], and not just sample inhomogeneity? All above simulations were made assuming particular $T_c(\sigma)$ dependence typical for aluminium: as smaller the cross section, as higher the $T_c$ (Fig. 3, insert). It means that for aluminium nanowires no feature below the bulk $T_c$ value ($T_c^{bulk} \sim 1.19$ K) can be accounted for geometry-dependent broadening. Very shallow $R(T)$ transitions stretched well below 1.2 K obtained on aluminium ultra-narrow nanowires [4,11] similar to Fig. 1 might really indicate presence of a new mechanism (e.g. quantum phase slip) having no trivial interpretation.

## 5. Conclusions

We have shown that size-dependent variation of critical temperature of a superconductor is extremely important for interpretation of fluctuation phenomena in quasi-1D systems. Though this size effect is known for decades, its origin still does not have a commonly accepted explanation. Using as an example a typical aluminum nanowire, we have shown that dominating contribution to broadening of experimentally observed $R(T)$ transition above the bulk critical value $T_c^{bulk}$ might originate from geometrical inhomogeneity and finite length of the sample. In various superconducting materials the dependence of critical temperature on characteristic dimension has different sign and value. Therefore, our results obtained for aluminium might not have a universal validity. However, many of currently studied nanostructures are fabricated from superconducting materials (or under such conditions?) when size dependence of the critical temperature is an experimental reality. With decrease of a wire diameter usually the technologically-determined inhomogeneity $\delta\sigma/\sigma$ increases. Hence, the inhomogeneity-broadened superconducting transition should be more pronounced in state-of-the-art narrow structures. The issue is of a vital importance for interpretation of data related to the role of thermal and quantum fluctuations in low-dimensional superconductors.

## Acknowledgements

The work was supported by the EU Commission FP6 NMP-3 project 505457-1 ULTRA-1D.

Figure captions.

Fig. 1. Resistance vs temperature for three samples obtained by progressive reduction of diameter of the same aluminium nanowire with length $L = 10$ μm. Measuring current and SPM measured cross sections are indicated in the figure.

Fig.2. Bottom panel: SPM image of a typical aluminum wire with 4 probes. Top panel: SPM measured variation of the wire cross section σ along the sample including the node regions. Inset: recalculated distribution of cross sections of the thinnest part of the sample (excluding the nodes).

Fig. 3. Resistance vs. temperature for aluminum nanowire from Fig. 2. Solid circles stand for experimentally measured $R(T)$ dependence. Open circles correspond to simulated $R(T)$ dependence based on Eq. (2) and empirical $T_c(\sigma)$ data (inset) without taking into consideration node regions. Diamonds are results of similar calculation with contribution of the node regions within the limits indicated by arrows in Fig. 2. Inset: empirical dependence of critical temperature $T_c$ on cross section $\sigma$ for aluminum nanowires. Line is guide for eyes.

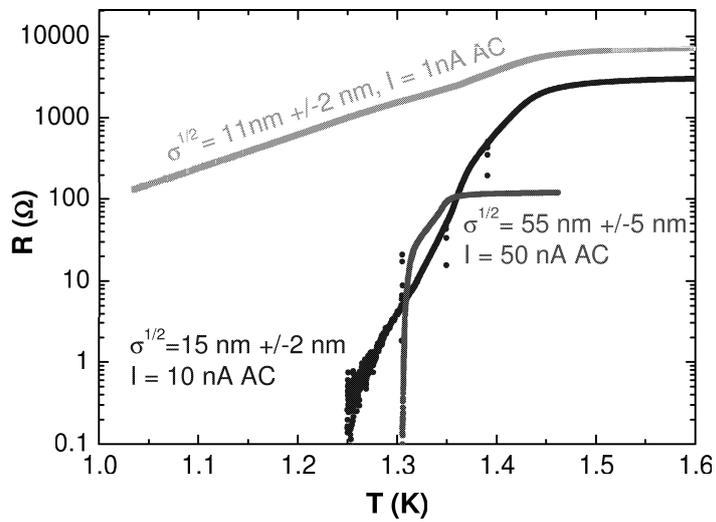

Fig. 1
Zgirski " Resistive state…"
Physica E
DECONS 06

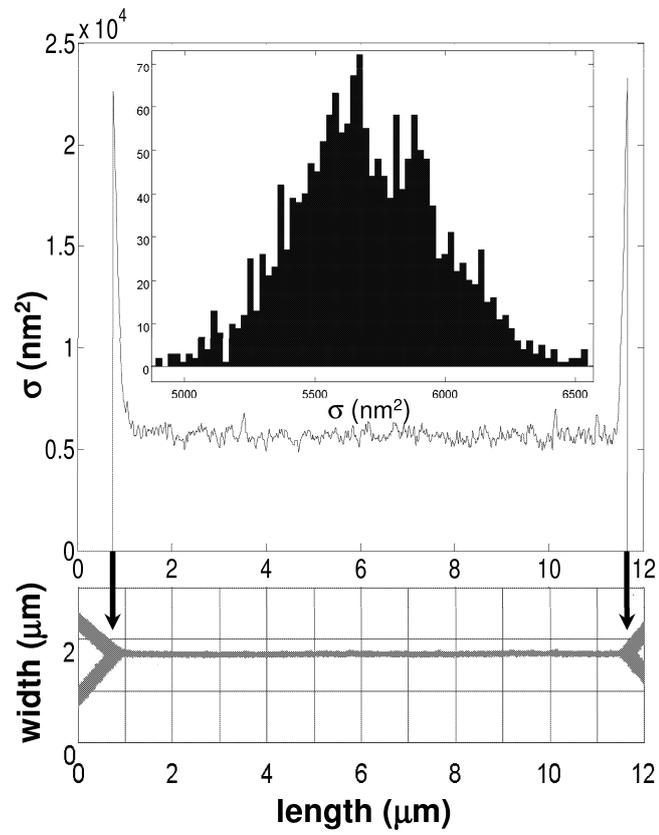

Fig. 2
Zgirski " Resistive state…"
Physica E
DECONS 06

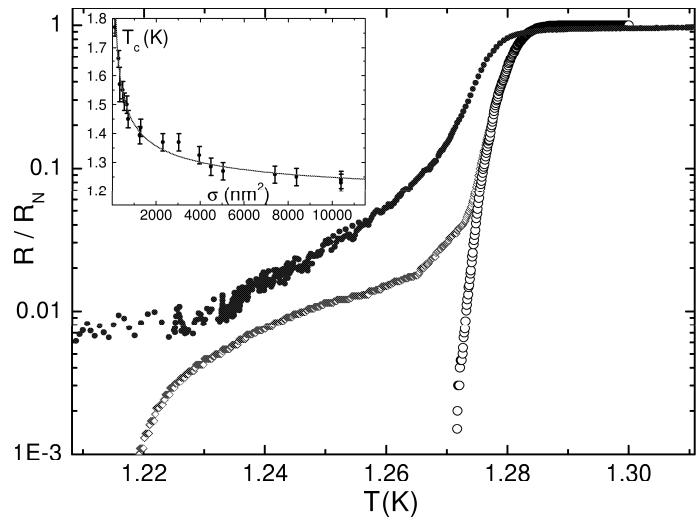

Fig. 3
Zgirski " Resistive state…"
Physica E
DECONS 06